\documentclass[prc,aps,floatfix,twocolumn,superscriptaddress,nofootinbib,showpacs,preprintnumbers,
amsmath,amssymb,amsfonts,widetable] {revtex4-1}
\usepackage{bm}
\usepackage{soul}
\usepackage{array}
\usepackage{lipsum}
\usepackage{relsize}
\usepackage{mathrsfs}
\usepackage{amssymb}
\usepackage{amsmath}
\usepackage{graphicx}
\usepackage[usenames]{color}
\usepackage{pslatex}
\usepackage{booktabs}
\usepackage{physics}
\usepackage{float}
\usepackage{tcolorbox}
\usepackage{ulem}

\newcommand{\expo}{\mathlarger{e}}
\newcommand{\rhozero}{\rho_{\raisebox{-1.0pt}{\tiny\!0}}}
\newcommand{\epszero}{\varepsilon_{\raisebox{-0.5pt}{\tiny 0}}}
\newcommand{\fF}{\mathlarger{f}_{\raisebox{-0.5pt}{\tiny F}}}
\newcommand{\fS}{\mathlarger{f}_{\raisebox{-0.5pt}{\!\tiny SF}}}

\newcommand{\Lagrange}[1]{\mathscr{L}_{\raisebox{-0.5pt}{\scriptsize{#1}}}}

\begin{document}

\title{Elucidating the role of the surface energy in density functional theory}
\author{Md Jafrul Islam}
\affiliation{Department of Physics, Florida State University, Tallahassee, FL 32306, USA}
\affiliation{Department of Physics, University of Dhaka, Dhaka 1000, Bangladesh}

\author{Athul Kunjipurayil}
\affiliation{Department of Physics, Florida State University, Tallahassee, FL 32306, USA}

\author{J. Piekarewicz}
\affiliation{Department of Physics, Florida State University, Tallahassee, FL 32306, USA}

\author{A. Volya}
\affiliation{Department of Physics, Florida State University, Tallahassee, FL 32306, USA}

% \author{Md Jafrul Islam, Athul Kunjipurayil, J. Piekarewicz, and A. Volya}
% \affiliation{Department of Physics, Florida State University, 
% Tallahassee, FL 32306, USA}
\date{\today}

\begin{abstract}
 The saturation of symmetric nuclear matter---reflected in the nearly constant interior density of heavy nuclei---is a defining 
 property of nuclear matter. Modern relativistic energy density functionals (EDFs) calibrated exclusively to the properties of 
 finite nuclei, make robust predictions with quantified uncertainties about the bulk properties of symmetric nuclear matter in 
 the vicinity of the saturation density. Following the same fitting protocol, nonrelativistic Skyrme EDFs systematically predict 
 higher saturation densities than their relativistic counterparts. To investigate this tension in the bulk limit, we study the 
 ground-state properties of hypothetical symmetric macroscopic nuclei containing thousands of nucleons. Using both 
 relativistic and non-relativistic EDF frameworks, we extract the corresponding liquid-drop parameters. We find a clear 
 correlation between the volume and surface energy coefficients: Skyrme models, which saturate at higher densities, develop 
 softer and more diffuse surfaces with lower surface energies, whereas relativistic EDFs, which saturate at lower densities, 
 produce more defined and less diffuse surfaces with higher surface energies. This compensating behavior allows both classes 
 of models to reproduce empirical nuclear radii despite their distinct saturation properties. Our analysis suggests that the apparent 
 disparity in saturation densities arises from the intrinsic balance among saturation density, bulk binding energy, and surface 
 tension, rather than from the fitting protocol.
\end{abstract}

%\pacs{25.30.Bf, 27.40.+z, 21.10.Gv, 21.10.Ft}
\maketitle

\section{Introduction}
\label{sec:introduction}
Nuclear saturation---the existence of an equilibrium density related to the nearly uniform interior density of heavy nuclei---is 
an important property of nuclear matter\,\cite{Horowitz:2020evx}. Symmetric Nuclear Matter (SNM), an idealized, infinite 
system at zero temperature with equal numbers of neutrons and protons interacting solely via the strong nuclear force, has 
traditionally served to elucidate the complex mechanisms responsible for the onset of saturation. Indeed, theoretical models 
calibrated by the ground state properties of finite nuclei predict that the energy per nucleon of SNM exhibits a minimum at a 
density of $\rhozero\!\approx\!0.15\,{\rm fm}^{-3}$. As such, for a nuclear Fermi system at zero temperature, the saturation 
density and the energy per nucleon at saturation ($E/A\!\approx\!-16\,{\rm MeV}$) are closely related to the central density 
of heavy nuclei\,\cite{Horowitz:2020evx} and the volume term of the semi-empirical mass 
formula\,\cite{Weizsacker:1935,Bethe:1936,Myers:1969}, respectively. 

A fully ab-initio description of nuclei based directly on QCD is not yet available. In chiral effective field theory (EFT), which 
provides a controlled low-energy framework, nuclear interactions are constructed by integrating out QCD degrees of freedom. 
Within this approach, nuclear saturation arises from a delicate interplay between kinetic and potential contributions, with significant 
influence from many-body effects such as three-nucleon forces that emerge because the underlying QCD dynamics is not resolved 
explicitly\,\cite{Hebeler:2010xb,Drischler:2017wtt,Hebeler:2020ocj}. Since current chiral EFT interactions still struggle to reproduce 
the empirical saturation point, insights from nuclear density functional theory (DFT) remain important for constraining these 
interactions\,\cite{Drischler:2024ebw}.

Within the framework of DFT, nonrelativistic energy-density functionals of Skyrme type rely on density-dependent effective
interactions to reproduce the correct saturation properties\,\cite{Vautherin:1971aw,Negele:1972zp,Negele:1981tw}. These 
density-dependent terms effectively encode contributions from three-nucleon and higher-body forces. Instead, in the relativistic 
case, nuclear saturation emerges from a sensitive interplay between the Lorentz structure of the scalar and time-like vector 
potentials\,\cite{Walecka:1974qa,Serot:1984ey,Serot:1997xg}. In the past,  the ``empirical" saturation point has served to calibrate 
the parameters of both non-relativistic and relativistic models. However, with the advent of powerful computers and machine learning 
algorithms, the optimization and calibration of energy density functionals is now free from any assumption on the bulk properties of 
infinite nuclear matter. That is, the calibration solely relies on genuine physical observables that can be measured in the laboratory. 
Hence, the saturation properties now emerge as predictions of the model with proper statistical uncertainties; see for 
example\,\cite{Kortelainen:2010hv,McDonnell:2015sja,Chen:2014sca,Chen:2014mza,Giuliani:2022yna}.

Given the central role of the saturation density in constraining nuclear dynamics, a natural question arises:  How well is the 
nuclear saturation point constrained within density functional theory? Since most models employ similar fitting protocols, one 
might naturally expect a high degree of agreement among them. Yet, as we demonstrate below, subtle correlations among bulk 
and surface properties can lead to systematic differences in the predicted saturation densities. Indeed, nonrelativistic Skyrme 
functionals systematically predict a higher saturation density as compared to their relativistic counterparts\,\cite{Drischler:2024ebw}. 
This fact alone would suggest that Skyrme functionals should predict smaller charge radii for heavy nuclei as compared to relativistic 
models. Yet both classes of models ``reproduce'' experimental charge radii, as these data were used in their calibration. Thus, we 
conjecture that in order to match experiment, relativistic models must compensate by having a stiffer surface energy than Skyrme 
models. Since infinite nuclear matter is insensitive to surface effects, this mismatch may be a key factor behind the discrepancy in 
the prediction of the saturation density. Understanding the surface properties of these two class of models, and the connection 
between saturation density and surface behavior within constraints set by empirical observations of finite nuclei, is the primary 
goal of this paper. This work is timely in light of the ``CREX-PREX dilemma''\,\cite{Reinhard:2022inh}, as surface contributions, 
particularly those associated to the spin-orbit interaction, may play a critical role in resolving the 
tension\,\cite{Yue:2024srj,Zhao:2024gjz,Kunjipurayil:2025xss}.

The manuscript is organized as follows. Section \ref{sec:formalism} reviews the formalism of both non-relativistic Skyrme and 
covariant energy density functionals. In Section \ref{sec:results}, we compute surface-energy coefficients for symmetric nuclear 
matter fitting a liquid-drop parametrization. The extraction of the surface energy is carried out by computing ground-state energies 
of artificial nuclei containing thousands of nucleons and with the Coulomb repulsion turned off. We offer a brief summary and our 
conclusions in Sec.~\ref{sec:conclusions}.  

%%%%%%%%%%%
\section{Formalism}
\label{sec:formalism}

\subsection{Non-Relativistic Skyrme Interaction}
\label{sec:SHF}
The phenomenological Skyrme effective interaction is at the foundation for self-consistent non-relativistic Hartree–Fock (HF) studies 
in nuclear structure. It is based on the idea that the energy functional associated with the short-range nuclear force can be expressed 
in terms of zero-range (contact) interactions organized as a momentum (gradient) expansion. In the language of effective field theory, 
this corresponds to a low-momentum expansion of short-range physics, akin to the contact terms in chiral or pionless EFT, but without 
an explicit power-counting or renormalization scheme to control order-by-order accuracy; and by contrast, a phenomenological density 
dependence is introduced to mimic medium effects rather than arising from a controlled EFT expansion. Because of the nature of the 
contact interaction, the Hartree and Fock contributions share the same local functional form, which simplifies the HF equations. This 
also significantly simplifies the evaluation of interaction matrix elements in a single-particle basis, reducing them to local integrals.

The standard Skyrme force defining the effective nucleon-nucleon interaction has the following 
form\,\cite{Vautherin:1971aw, Chabanat:1997un}:
\begin{align}
v_{12} &= t_0(1+x_0P_\sigma)\delta (\vec{r}_1-\vec{r}_2)+\frac{1}{2}t_1(1+x_1P_\sigma)\big[\delta (\vec{r}_1-\vec{r}_2)k^2 \nonumber\\
&+k'^2\delta (\vec{r}_1-\vec{r}_2)\big]+ t_2(1+x_2P_\sigma)\vec{k'}\cdot\delta (\vec{r}_1-\vec{r}_2)\vec{k} \nonumber \\
&+\frac{1}{6}t_3(1+x_3P_\sigma)\rho^{\alpha}\delta(\vec{r}_1-\vec{r}_2)\nonumber \\
&+iW_0(\vec{\sigma}_1+\vec{\sigma}_2)\cdot\vec{k'}\times\delta (\vec{r}_1-\vec{r}_2)\vec{k},
\label{eq: 2.14}
\end{align}
where, $\vec{k}=\frac{1}{2i}(\vec{\nabla}_1-\vec{\nabla}_2)$ and $\vec{k'}=-\frac{1}{2i}(\vec{\nabla}'_1-\vec{\nabla}'_2)$ are the relative 
momentum operators acting on the right and left respectively. $P_\sigma=\frac{1}{2}(1+\vec{\sigma_1}\cdot\vec{\sigma_2})$ is the 
spin-exchange operator. Finally, the parameters $t_0, t_1, t_2, t_3, x_0,x_1,x_2,x_3, W_0$, and $\alpha$ determine the strengths of the 
different interaction terms, and are fitted to nuclear data.

The $t_0$ term is a central, zero-range (momentum-independent, $s$-wave) contribution that largely sets the overall scale of the interaction 
and provides a leading contribution to the volume energy of nuclear matter. To avoid an unphysical collapse at high density, a 
density-dependent $t_3$ term is included; for typical choices of $\alpha>0$, it yields a repulsive contribution that grows with density and 
enables the saturation of symmetric nuclear matter. The momentum-dependent $t_1$ and $t_2$ terms emulate finite-range effects while 
remaining zero-range, thereby imparting realistic momentum dependence to the mean field (effective mass). These gradient terms also 
govern the surface energy and density diffuseness of finite nuclei. The parameter $W_0$ quantifies the strength of the phenomenological 
spin-orbit potential that accounts for the observed nuclear shell structure. In some models an additional parameter is introduced to incorporate 
isovector spin-orbit effects\,\cite{REINHARD1995467,Yue:2024srj}. Finally, a long-range Coulomb interaction is added to account for the 
repulsion among protons. 

In this work we employ five of the most representative Skyrme parameter sets, each calibrated to reproduce ground-state properties of 
finite nuclei and/or nuclear-matter bulk parameters. These parameter sets are commonly used within a self-consistent Skyrme–Hartree–Fock 
(SHF) framework; they are summarized in Table\,\ref{Table1}.

%%%%%  Table 0  %%%%%
\begin{widetext}
\begin{center}
\begin{table}[h]
\begin{tabular}{|l||c|c|c|c|c|c|c|c|c|c|}
\hline\rule{0pt}{2.5ex}   
\!\!Model   &  $t_0$    &  $t_1$    &  $t_2$    &  $t_3$    &  $x_0$       
            &  $x_1$    &  $x_2$    &   $x_3$   &  $\alpha$ &  $W_0$ \\
\hline
\hline
SKM*      & -2645.00 & 410.00  & -135.00 &  15595.00 & 0.090   & 0.000   &  0.000   & 0.000  & $1/6$ & 130.00  \\
RATP      & -2160.00 &  513.00 & 121.00  &  11600.00 & 0.418   & -0.360   & -2.290  & 0.586  & $1/5$ & 120.00 \\
SLy4       & -2488.91 &  486.82 & -546.39 &  13777.00 & 0.834   & -0.344   & -1.000  & 1.354  & $1/6$ & 123.00  \\ 
KDE0v1  & -2553.08 & 411.70  & -419.87 &  14603.61 & 0.648   & -0.3472  & -0.926  & 0.948  & 0.1673 & 124.41 \\
LNS        & -2484.97 &  266.74 & -337.14 & 14588.20 & 0.063   & 0.65845  & -0.954  & -0.034  & 0.1667 & 96.00 \\
\hline
\end{tabular}
\caption{Model parameters for the five Skyrme interactions used in this work. These are: 
SKM*\,\cite{BARTEL198279}, RATP\,\cite{rayet1982nuclear}, SLy4\,\cite{CHABANAT1998231}, KDE0v1\,\cite{Agrawal:2005ix}, 
and LNS\,\cite{Cao:2006}. The parameter $t_0$ is given in $\rm MeV\,fm^{3}$; $t_1$, $t_2$, and $W_0$ are given in $\rm MeV\,fm^5$; 
$t_3$ is given in $\rm MeV\,fm^{3(1+\alpha)}$; $x_i$'s and $\alpha$ are dimensionless.}
\label{Table1}
\end{table}
\end{center}
\end{widetext}
%%%%%%%%%%%%%%%%%%%%

%%%%%%%%%%%
\subsection{Covariant Energy Density Functional}
\label{sec:RMF}

The relativistic formalism based on a Lagrangian density that has undergone substantial refinements
throughout the years may be expressed as follows:
%%%%%%%%
\begin{equation}
 \Lagrange{} = \Lagrange{0} + \Lagrange{1} + \Lagrange{2}.
\end{equation}
%%%%%%%%
The first ``non-interacting'' term ($\Lagrange{0}$) consists of the kinetic energy of all the 
constituents, which (as shown below) includes an isodoublet nucleon field ($\psi$), the 
photon field ($A_{\mu}$) that accounts for the Coulomb repulsion, and three ``meson" fields 
responsible for mediating the nuclear interaction. In this work these are given by two 
isoscalar mesons: one scalar ($\phi$) responsible for the intermediate-range attraction 
and one vector ($V_{\mu}$) responsible for the short-range repulsion;  and a single 
isovector-vector field (${\bf b}_{\mu}$), that accounts for the isospin dependence of the 
nuclear interaction.

In turn, $\Lagrange{1}$ contains the Yukawa couplings of the various meson fields to the
corresponding scalar and vector bilinears in the nucleon field. 
That is\,\cite{Walecka:1974qa,Serot:1984ey,Serot:1997xg},
%%%%%%%%
%\begin{widetext}
\begin{eqnarray}
\Lagrange{1}=\bar\psi \left[g_{\rm s}\phi  \!-\! g_{\rm v}\gamma^{\;\mu}V_{\mu}
             \!-\! g_{\rho}\gamma^{\;\mu}\,{\bf b}_{\mu}\!\cdot\!\frac{\boldsymbol{\tau}}{2}
              \!-\! \frac{e}{2}\gamma^{\;\mu}A_{\mu}(1\!+\!\tau_{3})\right]\psi.
 \label{L1}
\end{eqnarray}
%\end{widetext}
%%%%%%%%
For example, the scalar-isoscalar field $\phi$ couples to the corresponding scalar-isoscalar density 
$\bar\psi\psi$ with strength  $g_{\rm s}$. Note that we use the standard (Weyl) representation of the 
Dirac gamma matrices\,\cite{Peskin1995} and that $\boldsymbol{\tau}$ is the vector containing the 
three Pauli matrices. 

Whereas the above Lagrangian density provided fundamental insights into the nuclear dynamics,
notably the natural emergence of nuclear saturation as a results of the Lorentz covariance manifested 
in the Lagrangian density\,\cite{Horowitz:1981xw}, several experimental features were difficult to
reproduce. To remedy these deficiencies a collection of both unmixed and mixed meson self-interactions
have been steadily incorporated into the following Lagrangian density\,\cite{Boguta:1977xi,Serot:1984ey,
Mueller:1996pm,Lalazissis:1996rd,Serot:1997xg,Horowitz:2000xj,Todd-Rutel:2005fa,Chen:2014sca,
Chen:2014mza,Salinas:2023qic}:
%%%%%%%%
\begin{widetext}
\begin{equation}
\Lagrange{2} =   - \frac{1}{3!} \kappa\,\Phi^3 - \frac{1}{4!} \lambda\Phi^4 
                            + \frac{1}{4!} \zeta (W_\mu W^\mu)^2 
                            + \Lambda_{\rm v} (\boldsymbol{B}_\mu \cdot \boldsymbol{B}^{\,\mu})(W_\mu W^\mu),
 \label{L2}
\end{equation}
\end{widetext}
%%%%%%%%
where $\Phi\!\equiv\!g_{\rm s}\phi$, $W_\mu\!\equiv\!g_{\rm v}V_\mu$, and 
$\boldsymbol{B}_\mu\!\equiv\!g_{\rho}\boldsymbol{b}_\mu $. Briefly, the cubic and quartic scalar terms 
($\kappa$ and $\lambda$) serve to soften the equation of state of symmetric nuclear matter around saturation 
density\,\cite{Boguta:1977xi}, the quartic vector term ($\zeta$) also softens the equation of state of symmetric 
nuclear matter, but at much higher densities. Finally, the mixed isoscalar-isovector term ($\Lambda_{\rm v}$)
was introduced in Ref.\,\cite{Horowitz:2000xj} to soften the density dependence of the symmetry energy. 
Including the scalar mass $m_{\rm s}$ that sets the range of the scalar interaction, the model includes a total 
of 8 free parameters. Model parameters calibrated to the ground-state properties of finite nuclei are listed in 
Table\,\ref{Table2}. Note that no derivative couplings appear in the formulation of the Lagrangian density. Rather,
gradient terms that determine the surface properties of the model, such as the strength of the spin-orbit 
interaction, are generated dynamically through the relation between upper and lower components of the
Dirac spinors. This is unlike conventional Skyrme parameterizations where gradient terms are directly
incorporated into the effective interaction. 

%%%%%  Table I  %%%%%
\begin{widetext}
\begin{center}
\begin{table}[h]
\begin{tabular}{|l||c|c|c|c|c|c|c|c|}
\hline\rule{0pt}{2.5ex}   
\!\!Model   &  $m_{\rm s}$    &  $g_{\rm s}^2$  &  $g_{\rm v}^2$  &  $g_{\rho}^2$  
                  &  $\kappa$       &  $\lambda$    &  $\zeta$       &   $\Lambda_{\rm v}$  \\
\hline
\hline
FSUGarnet    & 496.939 &   110.3492 & 187.6947 & 192.9274 & 3.2602  & $-$0.003551  & 0.0235  & 0.043377  \\
RMF022        & 497.5872 & 109.3475 & 185.5452 &  127.1648 & 3.1692  & $-$0.002786 & 0.0235  & 0.025964  \\
TFa          & 502.2000 &  106.5045 &176.1779 &  97.3556 & 3.1824   & $-$0.003470   & 0.0200  & 0.012670  \\
FSUGold2     & 497.479 &  108.0943 & 183.7893 &   80.4656 & 3.0029  & $-$0.000533  & 0.0256  & 0.000823  \\ 
NL3               & 508.194 &  104.3871 & 165.5854 &   79.6000 & 3.8599  & $-$0.015905  & 0.0000  & 0.000000  \\
\hline
\end{tabular}
\caption{Model parameters for the five covariant energy density functionals used in this work. These are: 
FSUGarnet and RMF022\,\cite{Chen:2014mza}, TFa\,\cite{Fattoyev:2013yaa}, FSUGold2\,\cite{Chen:2014sca}, 
and NL3\,\cite{Lalazissis:1996rd}. The parameter $\kappa$ and the scalar mass $m_{\rm s}$ are given in MeV. 
The masses of the $\omega$-meson, the $\rho$-meson, and the nucleon have been fixed near their experimental 
values at  $m_{\rm v}\!=\!782.5\,{\rm MeV}$, $m_{\rho}\!=\!763.0\,{\rm MeV}$, and $M\!=\!939.0\,{\rm MeV}$, respectively.}
\label{Table2}
\end{table}
\end{center}
\end{widetext}
%%%%%%%%%%%%%%%%%%%%

\subsection{Pairing Effect: Constant Gap Approach}
\label{sec:Pairing}
For our purposes, addressing ground-state properties of large hypothetical nuclei, a discussion of physics beyond Hartree–Fock, 
such as pairing correlations, is not central; the monopole component of pairing is already included in the 
interactions\,\cite{volya2002better}, and an explicit treatment of pairing correlations and the associated binding energy gain would 
amount to double counting\,\cite{Chen:2013jsa}. Nevertheless, pairing plays an important role for surface and shape properties 
because it slightly mixes the occupancies near the Fermi surface. This effect becomes paramount for the very large, hypothetical 
nuclear systems we consider: as the density of single-particle states grows, identifying an exact Fermi surface becomes impractical, 
and iterative self-consistent mean-field solutions tend to oscillate between different Slater determinants. The mixing provided by 
pairing stabilizes iterative solutions, and is also well known to smooth the density profile\,\cite{zelevinsky2003Nuclear}.

We therefore adopt a simple constant-gap treatment of pairing, following the BCS 
approach\,\cite{Reinhard1991,volya2002better,zelevinsky2003Nuclear,zelevinsky2004Pairing}, where the gap is taken from 
empirical systematics. A commonly used parametrization in this approximation is\,\cite{Reinhard1991,Seif2023}
\begin{equation}
  \Delta \;=\; \frac{11.2~\mathrm{MeV}}{\sqrt{A}}.
\end{equation}
This yields the standard BCS occupation probability for a single-particle state of energy $\epsilon$:
\begin{equation}
  w \;=\; \frac{1}{2}\!\left(
  1 - \frac{\epsilon - \epsilon_F}%
  {\sqrt{(\epsilon - \epsilon_F)^{2} + \Delta^{2}}}
  \right),
\end{equation}
where $\epsilon_F$ is the Fermi energy, adjusted to satisfy the particle-number constraint. These occupations are then used to obtain 
the self-consistent mean field. We emphasize that the stabilizing effect arises from the redistribution of occupations near the Fermi surface; 
any reasonable pairing-induced smoothing produces the same stabilization, so the precise value of $\Delta$ or a more elaborate pairing 
treatment does not affect our conclusions for the (hypothetically) very large systems considered here.

%%%%%%%%%%%%%%%%%%%%
\section{Results}
\label{sec:results}
The main objective of this paper is to compare nonrelativistic and relativistic predictions for the surface energy, which, through 
its connection to surface diffuseness, may help reconcile how models that reproduce rms radii can still predict different nuclear 
saturation densities. Our strategy is to employ artificially large (hypothetical) nuclei to sharpen the extraction of surface contributions 
by suppressing shell effects and curvature corrections, thereby approaching the infinite limit. Alongside the surface-energy coefficient, 
we also extract the volume-energy coefficient. Because these parameters are accessible 
through infinite-matter calculations, our discussion begins with those predictions before considering nuclei with very large mass 
number.

\subsection{Bulk properties of infinite nuclear matter}
\label{sec:INM}

Nuclear matter composed of protons and neutrons is characterized by the corresponding densities 
$\rho_{p}$ and $\rho_{n}$. For nuclear forces that are nearly isospin symmetric, it is convenient to 
discuss the special case of symmetric nuclear matter ($\rho_{p}=\rho_{n}$) and then examine small 
departures from this limit. Such deviations are commonly expressed in terms of the neutron--proton 
asymmetry $\alpha\!\equiv\!(\rho_{n}-\rho_{p})/(\rho_{n}+\rho_{p})$, and the resulting change in the 
energy per nucleon, $\mathlarger{\mathlarger{\varepsilon}}(\rho,\alpha)$, defines the symmetry energy. 
The symmetry energy, which quantifies the energy cost of converting protons into neutrons or vice versa, is 
defined as
%%%
\begin{equation}
 S(\rho) = \frac{1}{2}\!\left(\frac{\partial^{2}\mathlarger{\mathlarger{\varepsilon}}}
           {\partial\alpha^{2}}\right)_{\!\!\alpha=0}.
 \label{SymE}
\end{equation}
%%%

In the vicinity of the saturation density $\rhozero$, the density dependence of both the energy of symmetric nuclear 
matter as well as the symmetry energy are encoded in a few bulk parameters. That is,
%%%
\begin{subequations}
\begin{align}
 & \mathlarger{\varepsilon}_{{}_{\rm SNM}}(\rho) \equiv \mathlarger{\varepsilon}(\rho,\alpha\equiv0)
      = \epszero + \frac{1}{2}K_{0}\,x^{2}+\ldots \\
 & {S}(\rho) = J + L\,x + \ldots 
 %\frac{1}{2}K_{\rm sym}\,x^{2}\ldots   
\end{align} 
\label{EandS}
\end{subequations}
%%%
where $x\!=\!(\rho-\rhozero)\!/3\rhozero$ is a dimensionless parameter that quantifies the deviations 
of the density from its value at saturation. 

Predictions for these bulk parameters for both types of EDFs are listed in Table\,\ref{Table3}. We confirm that 
Skyrme-based nonrelativistic EDFs predict higher saturation densities than relativistic ones, with the LNS parametrization 
yielding the highest saturation density among the Skyrme sets considered. As noted by the authors\,\cite{Cao:2006}, this 
leads to comparatively higher central densities and, consequently, to the expectation of smaller radii relative to those 
predicted by relativistic models. Relativistic EDFs also tend to predict a higher binding energy per nucleon than their Skyrme 
counterparts. This pronounced difference at the predicted saturation points is clearly illustrated in Fig.\,1 of Ref.\,\cite{Drischler:2024ebw}, 
which includes an even broader set of models.

In addition, relativistic EDFs systematically predict a stiffer (more rapidly increasing) symmetry energy compared with Skyrme 
functionals. The inclusion of the $\Lambda_{\rm v}$ coupling in Eq.\,(\ref{L2}) was introduced precisely to soften the density 
dependence of the symmetry energy\,\cite{Horowitz:2000xj}. As shown in Table\,\ref{Table2}, and contrasted with the 
corresponding values of $L$ in Table\,\ref{Table3}, $\Lambda_{\rm v}$ plays a critical role in reducing the slope of 
the symmetry energy, $L$. All other things being equal, the larger the value of $\Lambda_{\rm v}$, the smaller the 
resulting value of $L$.

%%%%%%%%%%%%%%%%%%%%
\begin{center}
\begin{table}[h]
\begin{tabular}{|l||c|c|c|c|c|c|c|}
\hline\rule{0pt}{2.5ex}   
\!\!Model   &  $\rhozero$  &  $\epszero$  &  $K_{0}$  &  $J$  &  $L$ \\
\hline
\hline
SKM*\,\cite{BARTEL198279}       & 0.160 &  -15.78 & 216.70 & 30.03  & 45.78     \\
RATP\,\cite{rayet1982nuclear}    & 0.160 &  -16.00 & 240.00 & 29.27  & 32.39     \\
SLy4\,\cite{CHABANAT1998231}        & 0.160 &  -15.97 & 229.90 & 32.00  & 45.96     \\
KDE0v1\,\cite{Agrawal:2005ix}     & 0.165 &  -16.23 & 227.54 & 34.58  & 54.69     \\
LNS\,\cite{Cao:2006}     & 0.175 &  -15.32 & 210.85 & 33.43  & 61.45     \\
\hline
FSUGarnet\,\cite{Chen:2014mza}   & 0.153 &  -16.23 & 229.54 & 30.92  & 50.96     \\
RMF022\,\cite{Chen:2014mza}        & 0.152 &  -16.25 & 234.11 &  33.28 & 63.52      \\
TFa\,\cite{Fattoyev:2013yaa}           & 0.149 &  -16.23 & 245.12  & 35.05 & 82.50       \\
FSUGold2\,\cite{Chen:2014sca}      & 0.150 &  -16.27 & 237.88 &  37.59 & 112.72     \\ 
NL3\,\cite{Lalazissis:1996rd}            & 0.148 & -16.24 & 271.69 & 37.28  & 118.18     \\
\hline
\end{tabular}
\caption{Bulk parameters of infinite nuclear matter at saturation density 
$\rhozero$ as predicted by the non-relativistic Skyrme and relativistic
energy density functionals used in this work. The quantities $\epszero$ 
and $K_{0}$ represent the binding energy per nucleon and incompressibility 
coefficient of symmetric nuclear matter, whereas $J$ and $L$ denote the 
energy and slope of the symmetry energy---all evaluated at saturation. 
With the exception of the saturation density that is given in ${\rm fm}^{-3}$, 
all other quantities are expressed in MeV.}
\label{Table3}
\end{table}
\end{center}
%%%%%%%%%%%%%%%%%%%%

%%%%%%%%%%%%%%%%%%%%
\begin{figure}[ht]
\centering
\bigskip
\includegraphics[width=\linewidth]{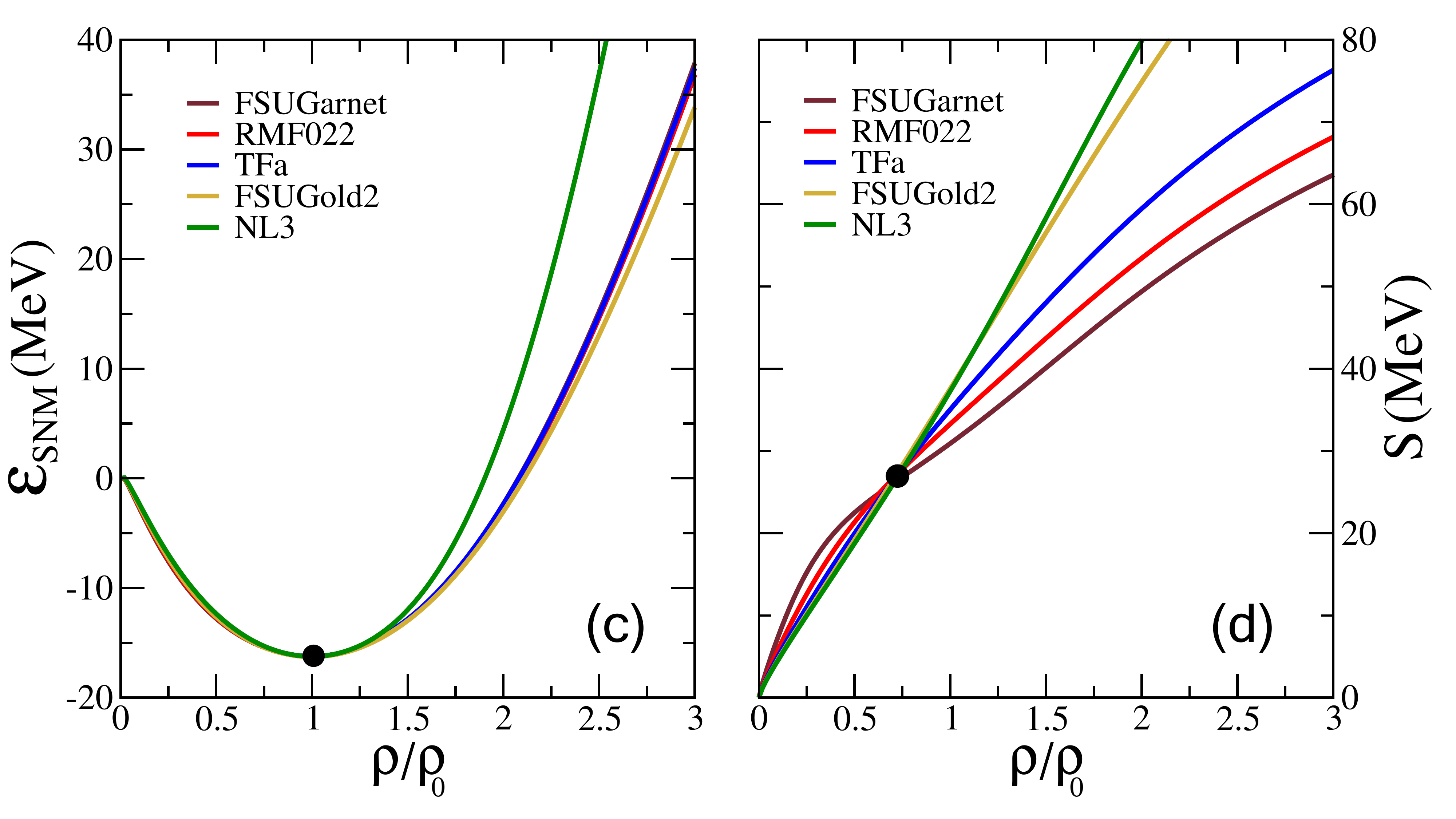}
\includegraphics[width=1.01\linewidth]{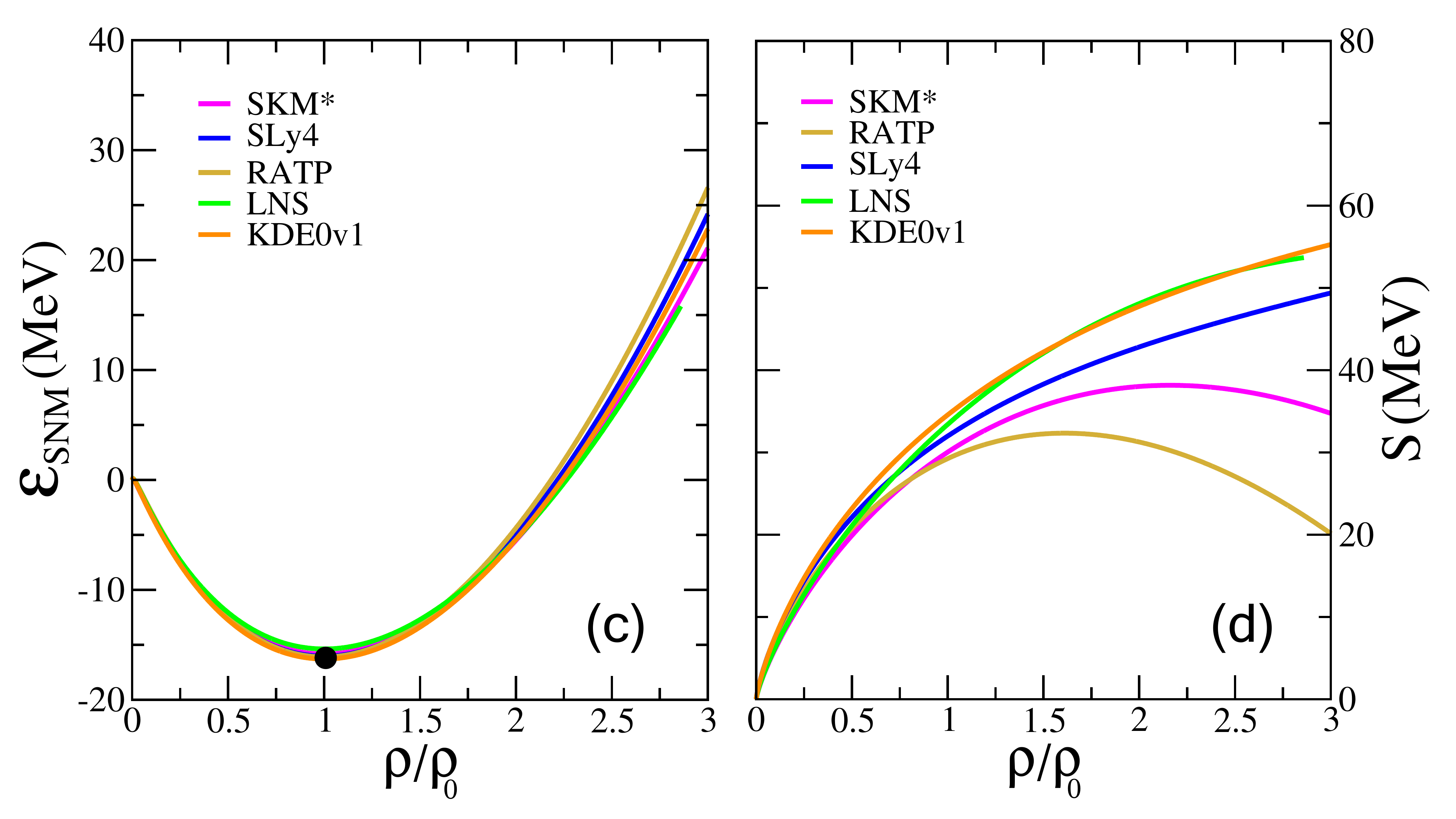}
\caption{(a) Binding energy per nucleon of symmetric nuclear matter as predicted by the 
five covariant energy density functionals used in this work. The small circle in the figure 
denotes the saturation point; see Table\,\ref{Table3}. (b) Predictions from the same five 
models for the density dependence of the symmetry energy. The small circle in the figure 
indicates that theoretical uncertainties in the value of the symmetry energy are minimized 
at a density $\rho\!\approx\!(2/3)\rhozero$. Panels (c) and (d) are the corresponding predictions
from the five Skyrme EDFs.}
\label{Fig1}
\end{figure}
%%%%%%%%%%%%%%%%%%%%

A more comprehensive view of the energy per nucleon in symmetric nuclear matter and of the symmetry energy, both displayed 
as functions of density, is presented in Fig.\,\ref{Fig1}. The top panels correspond to relativistic energy-density functionals, whereas 
the bottom panels show nonrelativistic Skyrme EDFs. The left-hand panels of Fig.\,\ref{Fig1} indicate that the binding energy per 
nucleon in symmetric nuclear matter is well constrained by the nuclear masses and charge radii used to guide the calibration 
procedure; at least in the vicinity of the saturation density. The outlying behavior of the NL3 model, characterized by an incompressibility 
coefficient (“curvature”) significantly larger than that of the other models (see Table\,\ref{Table3}), has been ruled out by measurements 
of the isoscalar giant monopole resonance~\cite{Garg:2018uam}.

In contrast, the symmetry energy shown on the right-hand panels is poorly determined, even at saturation density. This reflects the 
relatively small neutron-proton asymmetry of the nuclei used in the calibration procedure. The symmetry energy predicted by relativistic 
EDFs at a density of approximately $\rho\!\approx\!(2/3)\rho_{0}$ is relatively well constrained by experimental data on nuclear masses 
and charge radii\,\cite{Furnstahl:2001un}. In contrast, for Skyrme-type models the value near $\rho\!\approx\!(2/3)\rho_{0}$ is not as tightly 
constrained.

Unlike their relativistic counterparts, some Skyrme parameterizations predict a symmetry energy that decreases at densities above 
$\rho_{0}$, a trend that is difficult to justify\,\cite{LI2008113}. This behavior likely reflects the absence of rigorous constraints on the 
symmetry energy at suprasaturation densities, either from terrestrial nuclear experiments or from astrophysical observations of neutron 
stars. Since our present interest lies in nuclear properties near saturation density, such high-density behavior is not directly relevant to 
the present work. Looking ahead, the commissioning of new rare-isotope facilities, together with current and planned measurements 
of neutron skin thicknesses in neutron-rich nuclei\,\cite{Abrahamyan:2012gp,Adhikari:2021phr,Adhikari:2022kgg}, is expected to 
play a decisive role in constraining the density dependence of the symmetry energy.

%%%%%%%%%%%%%%%%%%%%
\subsection{Energy systematics for symmetric nuclei}
\label{sec:EnergySN}

Next, we focus on the extraction of the surface-energy coefficient. Numerous studies, both nonrelativistic and relativistic, have 
addressed this problem within the framework of the semi-infinite nuclear matter approximation; see, for example, 
Refs.\,\cite{Brack:1985vp,Treiner1986,DelEstal:1998ac}. In this approach, one considers a slab of nuclear matter at saturation 
density occupying half-space, with the other half remaining empty. This configuration naturally generates a surface at the interface, 
enabling one to study the energetic cost (per unit area) associated with creating such a boundary. However, the approach requires 
a careful, self-consistent treatment of the density profile and its surface diffuseness, over which the density varies. Establishing the 
minimization and self-consistency can be nontrivial and may require semiclassical methods, such as the Thomas–Fermi approach.

Alternatively—and this is the approach adopted here—the surface-energy coefficient, together with other bulk 
parameters of the liquid-drop model, may be extracted by following a leptodermous expansion, as outlined in 
Ref.\,\cite{Reinhard:2005nj}. Using a liquid-drop parametrization, the energy per nucleon can be expressed as
\begin{equation}
  \mathlarger{\varepsilon} = a_{\rm v} + a_{\rm s}\,A^{-1/3} + a_{\rm c}\frac{Z^{2}}{A^{4/3}} + a_{\rm sym}\frac{(N-Z)^{2}}{A^{2}}.
  \label{LDM}
\end{equation}
%%%
Here, $a_{\rm v}$, $a_{\rm s}$, $a_{\rm c}$, and $a_{\rm sym}$ denote the volume, surface, Coulomb, and asymmetry 
coefficients, respectively. Thus, when the energy is plotted as a function of $A^{-1/3}$ for a sequence of finite 
nuclei with equal numbers of protons and neutrons and with the Coulomb interaction turned off, the result should be 
a straight line with intercept $a_{\rm v}$ and slope $a_{\rm s}$. In particular, the value of $a_{\rm v}$ should closely 
match the energy per nucleon of symmetric nuclear matter at saturation density. 

%%%%%%%
\begin{center}
\begin{figure}[h]
\centering
\bigskip
\includegraphics[width=0.5\textwidth]{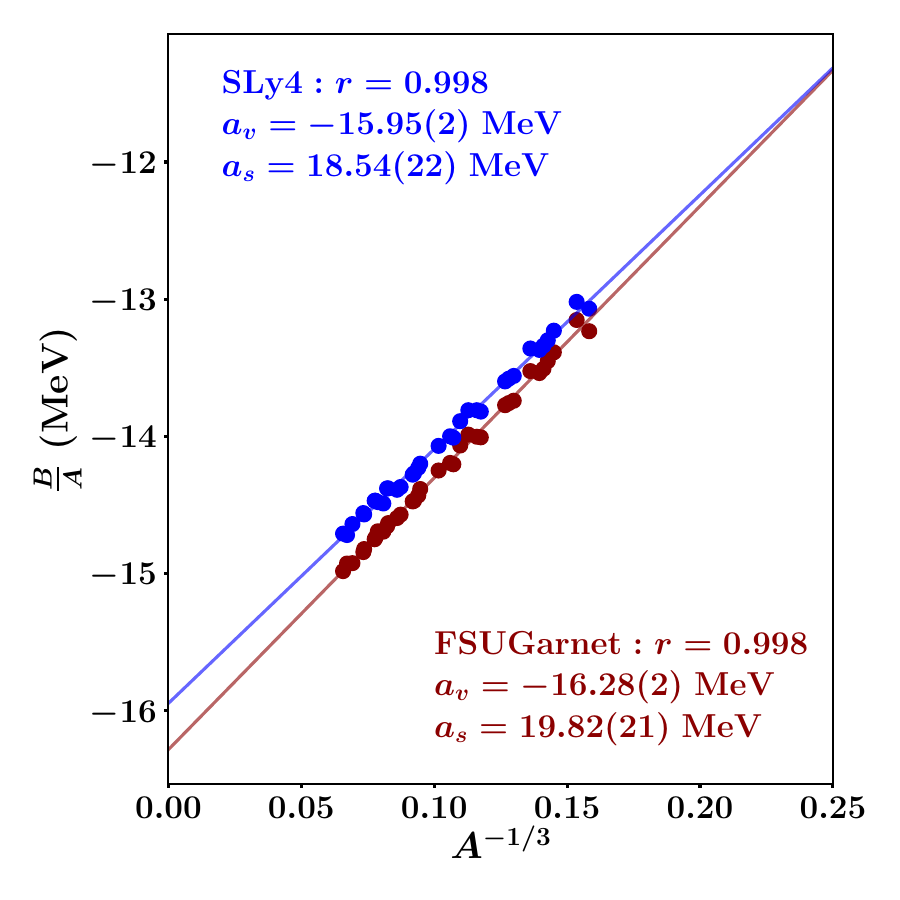}
\caption{Energy per nucleon for a collection of symmetric nuclei (without Coulomb) fitted to a semi-empirical mass 
formula $ \mathlarger{\varepsilon}=a_{\rm v}+a_{\rm s}A^{-1/3}$. Predictions are displayed with both Skyrme and 
FSUGarnet energy density functionals.}
\label{Fig2}
\end{figure}
\end{center}
%%%%%%%

%%%%%%%%%%%%%%%%%%%%
\begin{center}
\begin{table}[h]
\begin{tabular}{|l|c|c||l|c|c|}
\hline\rule{0pt}{2.5ex}   
\!\!Skyrme   &  $a_{\rm v}$  &  $a_{\rm s}$  &  RMF  &  $a_{\rm v}$  &  $a_{\rm s}$ \\
\hline
\hline
SKM*        &  -15.74  &  17.74    & FSUGarnet  &  -16.28  & 19.82 \\
RATP        &  -16.05  & 19.45    & RMF022      &  -16.31  & 19.95 \\
SLy4         &  -15.95  & 18.54    & 
TFa        &  -16.29  & 20.00 \\
KDE0v1    & - 16.20  & 17.88    & FSUGold2   &  -16.32  & 20.12 \\
LNS          & - 15.26  & 15.36    & NL3             &  -16.29  & 20.05 \\
\hline
\end{tabular}
\caption{Liquid-drop parameters extracted from a leptodermous expansion using a set of symmetric 
              nuclei spanning the range $A$ between about 250 and 3500 nucleons. The results illustrate how the development 
              of a surface imposes a stiffer energetic penalty in the relativistic models compared with their 
              nonrelativistic counterparts.}
\label{Table4}
\end{table}
\end{center}
%%%%%%%%%%%%%%%%%%%%

%%%%%%%%%%%%%%%%%%%%%%%%%%%%%%
\begin{figure*}[ht]
\centering
\begin{minipage}{0.45\textwidth}
    \centering
    \includegraphics[width=\linewidth]{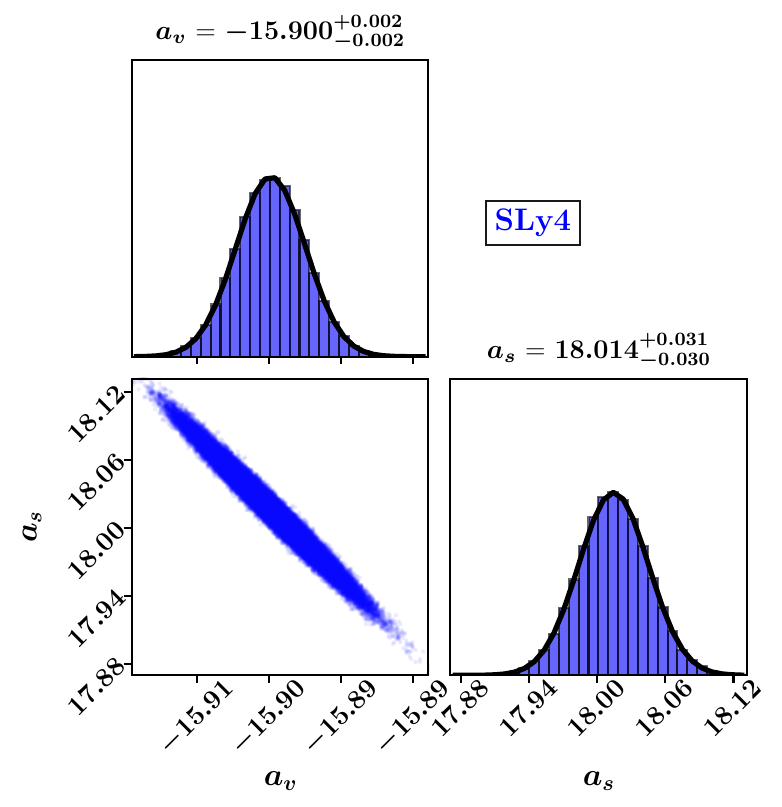}
    % \subcaption{}
    \label{fig:SLY4}
\end{minipage}
\hspace{2pt}
\begin{minipage}{0.45\textwidth}
    \centering
    \includegraphics[width=\linewidth]{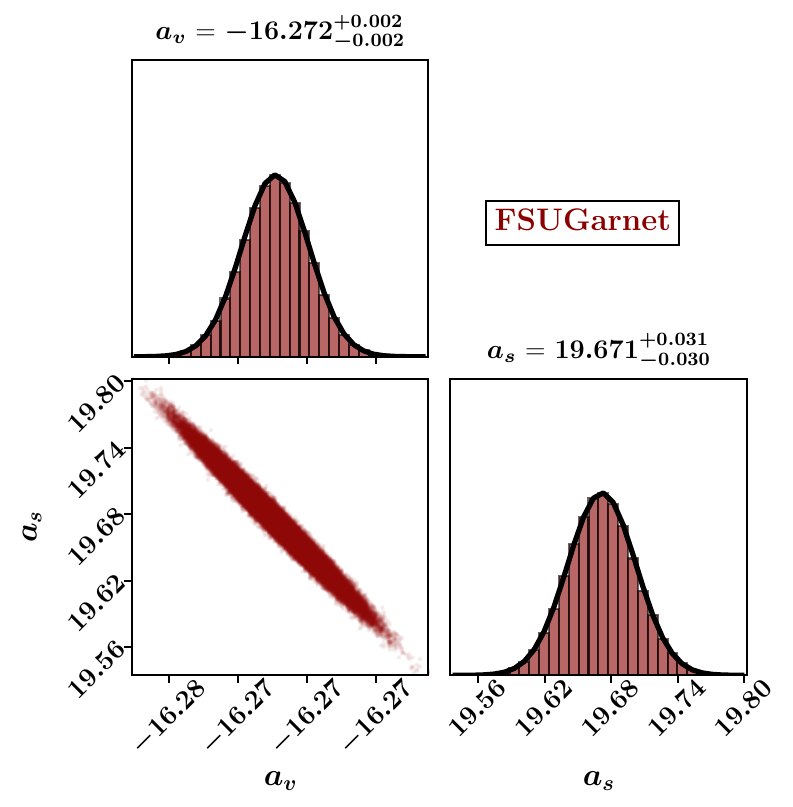}
    % \subcaption{}
    \label{fig:FSUGarnet}
\end{minipage}
\caption{Corner plot displaying the probability distribution and correlation coefficients for three 
              empirical parameters of the liquid-drop formula as obtained via a Metropolis-Monte-Carlo 
              method. The solid line and the associated labels represent the results obtained assuming 
              a normal distribution. Predictions are displayed with the SLy4 (in blue) and FSUGarnet 
              (in garnet) energy density functionals.}
\label{Fig3}
\end{figure*}
%%%%%%%%%%%%%%%%%%%%%%%%%%%%%%

Indeed, as shown in Fig.\,\ref{Fig2}, this is precisely the case. Using the FSUGarnet and SLy4 parametrizations for a 
set of 34 symmetric nuclei spanning the range of $A$ between about 250 and 3500 nucleons, a linear fit to the 
FSUGarnet predictions yields a volume term of $a_{\rm v}\!=\!-16.28(2)\,{\rm MeV}$, a surface-energy coefficient of 
$a_{\rm s}\!=\!19.82(21)\,{\rm MeV}$, and an almost perfect linear correlation characterized by a Pearson coefficient of 
$r\!=\!0.998$. The corresponding values for SLy4 are $a_{\rm v}\!=\!-15.95(2)\,{\rm MeV}$, 
$a_{\rm s}\!=\!18.54(22)\,{\rm MeV}$, and $r\!=\!0.998$. Our results are also summarized in Table\,\ref{Table4} for all 
EDFs considered in this work. Note that the volume terms extracted from the leptodermous expansion are largely consistent 
with the corresponding values of the binding energy per nucleon of symmetric nuclear matter at saturation density 
($\varepsilon_{0}$) listed in Table~\ref{Table3}. Moreover, the trend discussed earlier—namely, that the surface-energy 
coefficients predicted by  Skyrme functionals are systematically smaller than those obtained from relativistic mean-field 
models—is also confirmed by the results in Table\,\ref{Table4}. It is also worth noting that in relativistic models the ratio 
$a_{\rm s}/a_{\rm v}$ remains nearly constant at about 1.22, while in nonrelativistic models it ranges from 1.0 to 1.21, 
highlighting differences in the equilibrium surface density profile and diffuseness to be discussed in Sec.\,\ref{sec:Evolution}.

To further formalize the correlation between the volume and surface energy terms, we adopt the methodology described in 
Ref.\,\cite{Piekarewicz:2014kza}, where we consider a likelihood function constructed from an objective (or cost) function
$\chi^2$, defined as the sum of squared deviations between the empirical data for binding energy and the predictions of the liquid-drop model, and with a constant theoretical uncertainty. While the average value of the volume and surface coefficients are insensitive to the choice of  theoretical error, the standard deviation scales linearly with it, so we adopt the 3.8 MeV uncertainty suggested in Ref.\,\cite{Dobaczewski:2014jga}. Once the likelihood function is defined, we sample the parameter space using a standard Markov Chain Monte Carlo (MCMC) technique. The resulting probability distribution is shown in Fig.\,\ref{Fig3} for both the nonrelativistic SLy4  model and for the relativistic FSUGarnet model. 

Figure\,\ref{Fig3} shows strong correlation between the volume and surface coefficients $(a_{\rm v},a_{\rm s})$, the proportionality 
between the surface and volume terms reflects their common origin in the short-range nuclear interaction and the saturation property 
of nuclear matter, which together yield a nearly universal surface diffuseness. This correlation reflects the established interplay
between the saturation density, bulk binding energy, and surface tension that preserves the observed binding and radius of finite 
nuclei; see Fig.\,1 in Ref.\,\cite{Drischler:2024ebw}. Such an interplay serves as a natural starting point for discussing in the 
following section how the saturation properties influence the surface density profile and its diffuseness.

%%%%%%%%%%%%%%%%%%%%
\subsection{Evolution of nuclear radii}
\label{sec:Evolution}

\begin{figure}[t]
\centering
\bigskip
\includegraphics[width=\linewidth]{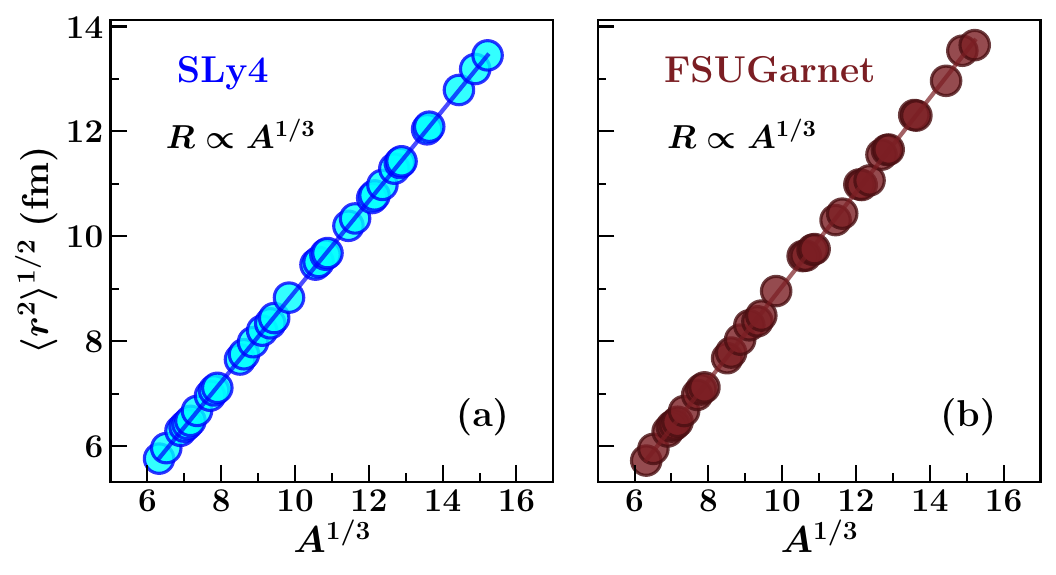}
% \vskip 1em
\caption{The characteristic $A^{1/3}$ scaling of nuclear radii, illustrating a direct manifestation of nuclear saturation. 
              Panels (a) and (b) display the results obtained with the SLy4 and FSUGarnet functionals, respectively, for a
              collection of symmetric nuclei spanning the $A\!=\!250$-$3500$ range.}
\label{Fig4}
\end{figure}

We begin the discussion of nuclear radii and surface diffuseness by highlighting the emergence of saturation for both real and 
hypothetical nuclei. To start, we display in Fig.\ref{Fig4} the dependence of the mean-square radius on $A$, where the 
nearly constant interior density suggests the following scaling: $R\!\equiv\!\langle r^{2}\rangle^{1/2}\propto A^{1/3}$.

%%%%%%%%%%
%%%%%%%%%%

In the previous section, we identified a strong correlation between the volume and surface energy coefficients---a feature 
that likely contributes to the model degeneracy where distinct saturation densities yield nearly identical rms radii. To further 
explore the underlying physics, we now examine how nuclear radii evolve with the surface-to-volume ratio, focusing on the 
density profile geometry and its associated surface diffuseness.

To formalize this idea, we invoke the symmetrized Fermi function. Although practically indistinguishable from the standard 
two-parameter Fermi form, the symmetrized version exhibits significantly improved analytic properties; see 
Ref.\,\cite{Sprung:1997} and references contained therein. In terms of the conventional two-parameter Fermi distribution, 
defined as
%%%%%%%%%
\begin{equation}
\fF(r) = \frac{1}{1+\expo^{(r-c)/a}},
\end{equation}
%%%
the symmetrized Fermi function is given by
%%%
\begin{equation}
\fS(r) \equiv \fF(r) + \fF(-r) - 1 =
\frac{\sinh(c/a)}{\cosh(r/a)+\cosh(c/a)},
\end{equation}
%%%%%%%%%
where $c$ denotes the half-density radius and $a$ the surface diffuseness. A major advantage of the symmetrized 
Fermi function over the conventional form is that its form factor, namely, its Fourier transform, can be evaluated in
closed analytic form\,\cite{Sprung:1997}. Consequently, all spatial moments of the distribution, such as the 
root-mean-square (rms) radius, can be computed exactly. The normalization (volume) term, as well as the first two 
moments of the spatial distribution are given by\,\cite{Piekarewicz:2016vbn}
%%%
\begin{subequations}
\begin{align}
4\pi &\int_0^\infty r^2 \fS(r)\,dr  
   = \frac{4\pi}{3}c\left(c^2 + \pi^2 a^2\right)\,,\\
R^2 & \equiv \langle r^{2} \rangle  = \frac{3}{5}c^{2} + 
  \frac{7}{5}(\pi a)^{2} \,,\\
 \langle r^{4} \rangle & = \frac{3}{7}c^{4} + 
  \frac{18}{7}(\pi a)^{2}c^{2} + \frac{31}{7}(\pi a)^{4}  \,.
\end{align}
 \label{SFMoments}
\end{subequations}
%%%%%%%%%

The interplay between the volume and surface terms in determining the rms radius is encapsulated in Eq.\,(\ref{SFMoments}). 
Reformulating Eq.\,(\ref{SFMoments}) to express the mass number $A$ in terms of the central density $\rho_{0}$, the rms radius 
$R$, and retaining terms up to second order in $\pi\,a/R$, we obtain
%%%%%%%%%%%%%%%%
\begin{equation}
 A = \frac{20}{9}\sqrt{\frac{5}{3}}\,\pi\, \rho_{0}\,R^{3} 
 \left(1 - \frac{3}{2}\frac{\pi^{2}a^{2}}{R^{2}}\right).
\label{eq:AofR}
\end{equation}
%%%%%%%%%%%%%%%%

In Skyrme models, which generally predict higher saturation densities than their relativistic counterparts (see Table\,\ref{Table3}), 
the corresponding half-density radius $c$ is expected to be smaller. This reduction can be compensated by a larger surface 
diffuseness $a$, which would be required to keep $A$ constant in Eq.\,\eqref{eq:AofR}. It is precisely this delicate balance that 
allows both classes of energy density functionals to reproduce the experimental charge radius of ${}^{208}$Pb, even though they 
differ in their predictions for the saturation density. 

To test this assertion, we have computed the root-mean-square (rms) baryonic radii--derived from the total 
baryon (cumulative proton and neutron) density--for a set of symmetric nuclei with $N\!=\!Z\!=\!A/2$ and with 
the Coulomb interaction switched off. By focusing on isospin-symmetric systems, contributions from the 
asymmetry term vanish, thereby enabling a clean isolation of the volume and surface contributions to the 
liquid-drop formula. 

In Table\,\ref{Table6}, we present a systematic study of baryonic radii for symmetric nuclei ranging 
from $Z\!=\!N\!=\!20$ up to $Z\!=\!N\!=\!750$, as predicted by FSUGarnet\,\cite{Chen:2014mza} and 
SLy4\,\cite{Chabanat:1997un}. The two horizontal lines in Table\,\ref{Table6}---above and below 
$Z\!=\!82$---delineate the transition from nuclei whose rms radii differences are dominated by 
the nuclear surface to larger systems that are instead dominated by the nuclear volume, where surface 
effects can no longer compensate for the different saturation densities predicted by the two models.

%%%%%%%%%%%%%%%%%%%%%%%%%%%%%%%%%%%%%%%
\begin{center}
\begin{table}[h]
\begin{tabular}{|c||c|c|c||c|c|c|}
\hline
\multicolumn{1}{|c||}{\rule{0pt}{2.5ex}} & 
\multicolumn{3}{c||}{\textbf{FSUGarnet}} &
\multicolumn{3}{c|}{\textbf{SLy4}} \\
\hline\rule{0pt}{2.5ex} 
\!\!Z  &  $R$  &  $c$  &  $a$ &  $R$  &  $c$  &  $a$ \\
\hline\rule{0pt}{2.5ex} 
\!\!20  &  3.277 & 3.556 & 0.478 & 3.359 & 3.680 & 0.478 \\
28  &   3.539 & 4.037 & 0.446 & 3.633 & 4.149 & 0.456 \\
40  &   4.028 & 4.598 & 0.506 & 4.069 & 4.684 & 0.496 \\
50  &   4.253 & 5.038 & 0.455 & 4.295 & 5.087 & 0.459 \\
\hline
82  &   4.983 & 6.009 & 0.479 & 4.989 & 6.013 & 0.481 \\
\hline
126 &  5.719 & 6.961 & 0.513 & 5.702 & 6.939 & 0.512 \\
184 &  6.458 & 7.903 & 0.554 & 6.426 & 7.867 & 0.548 \\
228 &  6.973 & 8.725 & 0.463 & 6.908 & 8.625 & 0.472 \\
308 &  7.673 & 9.617 & 0.494 & 7.602 & 9.514 & 0.502 \\
378 &  8.309 & 10.507 & 0.450 & 8.159 & 10.297 & 0.463 \\
476 &  8.957 & 11.366 & 0.443 & 8.791 & 11.138 & 0.454 \\
644 &  9.747 & 12.307 & 0.546 & 9.648 & 12.170 & 0.553 \\
750 & 10.304 & 13.060 & 0.527 & 10.163 & 12.871 & 0.530 \\
\hline
\end{tabular}
\caption{Root-mean-square baryon radius $R$, half-density radius $c$, 
and surface diffuseness $a$ for nuclei calculated with FSUGarnet and
SLy4; all quantities are in given in fm.}
\label{Table6}
\end{table}
\end{center}
%%%%%%%%%%%%%%%%%%%%%%%%%%%%%%%%%%%%%%%

For nuclei below $Z\!=\!82$, all radii are larger in SLy4---the model characterized by both a higher 
saturation density and a larger surface diffuseness. In contrast, for nuclei above $Z\!=\!126$, volume 
effects become dominant, leading to the larger radii predicted by FSUGarnet. Although SLy4 continues 
to exhibit a larger surface diffuseness for heavier nuclei, the larger half-density radius predicted by 
FSUGarnet more than compensates for the moderate increase in surface thickness. This crossover 
behavior provides further evidence of the strong correlation between the saturation density and 
the surface energy coefficient, underscoring how variations in $\rhozero$ and $a_{\rm s}$ may be
tuned to reproduce nuclear radii across different theoretical frameworks. 

In principle, the combination of higher saturation density and softer surface in Skyrme functionals can yield 
radii comparable to those of RMF models, which saturate at lower density but possess stiffer surface energy,
at least in the vicinity of $A\sim\ 200$. This compensating behavior highlights the delicate balance between 
bulk and surface contributions in finite nuclei. Thus, despite their differing microscopic foundations, both 
relativistic and nonrelativistic EDFs reproduce similar charge radii through offsetting trends in saturation 
density and surface energy.

\begin{figure}[ht]
\centering
\bigskip
\includegraphics[width=\linewidth]{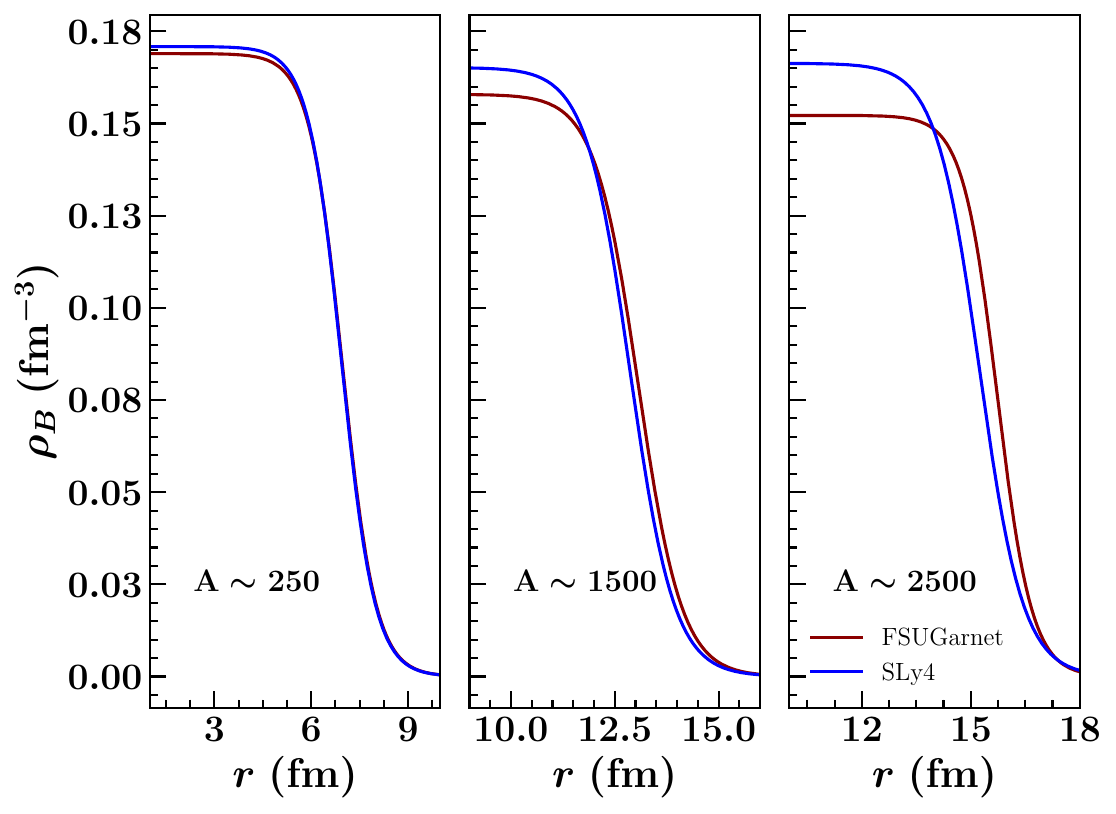}
\caption{Baryon density as predicted by FSUGarnet and SLy4 for hypothetical symmetric nuclei with 
              $A\!=\!252, 1500, 2504$.}
\label{Fig5}
\end{figure}

We conclude this section by illustrating this behavior in Fig.\,\ref{Fig5} by displaying baryon densities for 
hypothetical symmetric nuclei containing $252, 1500, 2504$ nucleons. 

%%%%%%%%%%%%%
%%%%%%%%%%%%%

%%%%%%%%%%%%%
\section{Conclusions}
\label{sec:conclusions}

In this work, we have investigated the interplay between the saturation density, bulk binding energy, and 
surface energy in both nonrelativistic and relativistic energy density functionals. By computing the binding 
energies and radii of a large set of symmetric nuclei---spanning mass numbers up to 
$A\!\sim\!3500$ nucleons, we have extracted the associated liquid-drop parameters through a systematic 
leptodermous expansion. We found a robust correlation between the volume and surface energy coefficients. 
Furthermore, Skyrme functionals predict higher saturation densities and a softer surface energy, whereas relativistic 
mean-field models saturate at lower densities but display a significantly stiffer surface energy. This compensating 
behavior ensures that both classes of models reproduce experimental charge radii despite their differing 
microscopic origins.

By exploring the evolution of nuclear radii across a wide mass range, we find that the influence of the 
surface contribution decreases markedly near $Z\!=\!82$. Hence, for lighter nuclei, the larger surface 
diffuseness associated with Skyrme functionals impacts the radius systematics, whereas in RMF models 
the lower saturation density is compensated by a smaller surface diffuseness, thereby yielding consistent radii. 
These results suggest that the apparent discrepancy in the saturation densities between Skyrme and relativistic EDFs arises not necessarily from deficiencies in the fitting protocol, but rather from the intrinsic structure of the functionals themselves. 
The saturation point of nuclear matter thus reflects a delicate 
balance between competing bulk properties---a balance that future microscopic EDFs should elucidate 
given their critical role in guiding the calibration of chiral interactions.

Finally, the present study lays the groundwork for a unified mapping between the parameters of relativistic 
and nonrelativistic functionals. In a forthcoming publication, we will exploit the fact that, in both sets of models, 
various bulk properties of infinite nuclear matter can be expressed directly in terms of the underlying model 
parameters. Consequently, properties such as the binding energy per nucleon, incompressibility, and symmetry 
energy among others, may serve as a bridge for connecting and constraining model parameters across 
different theoretical frameworks.

\bibliography{./ReferencesJP}

\begin{acknowledgments}\vspace{-10pt}
This material is based upon work supported by the U.S. Department of Energy Office of Science, 
Office of Nuclear Physics under Award Numbers DE-FG02-92ER40750 and DE-SC0009883.
\end{acknowledgments}

\end{document}